# Deep-learning recognition and tracking of individual nanotubes in low-contrast microscopy videos

Vladimir Pimonov[1], Said Tahir[1], Vincent Jourdain[1]

Affiliations:

1 Laboratoire Charles Coulomb (L2C), UMR 5221 CNRS-Université de Montpellier, Place Bataillon, Montpellier, FR-34095, France.

*Correspondence to: vincent.jourdain@umontpellier, vladimir.v.pimonov@gmail.com

## Abstract

This study addresses the challenge of analyzing the growth kinetics of carbon nanotubes using *in-situ* homodyne polarization microscopy (HPM) by developing an automated deep learning (DL) approach. A Mask-RCNN architecture, enhanced with a ResNet-50 backbone, was employed to recognize and track individual nanotubes in microscopy videos, significantly improving the efficiency and reproducibility of kinetic data extraction. The method involves a series of video processing steps to enhance contrast and used differential treatment techniques to manage low signal and fast kinetics. The DL model demonstrates consistency with manual measurements and increased throughput, laying the foundation for statistical studies of nanotube growth. The approach can be adapted for other types of *in-situ* microscopy studies, emphasizing the importance of automation in high-throughput data acquisition for research on individual nano-objects.

## Introduction

Carbon nanotubes (CNTs), discovered over three decades ago, continue to present unresolved questions and challenges. Their exceptional properties, both theoretically (*1, 2*) and experimentally demonstrated (*3*), make them desirable for electronic and optical devices. However, the widespread application of CNTs is hindered by the lack of control over their structure during growth. Therefore, developing highly selective synthesis methods is crucial for advancing CNT-based devices. This requires a deep understanding of the relationship between nanotube structure and selectivity, particularly kinetic selectivity. To address this, we developed a method based on *in situ* homodyne polarization microscopy (HPM), which is highly sensitive and can detect changes in optical absorption caused by a single carbon nanotube. The technique allows imaging tens to hundreds of individual carbon nanotubes during growth at up to 40 frames per second (*4*). However, the vast amount of information generated requires meticulous and time-consuming analysis to extract kinetic data.

This challenge, common in imaging-related fields, can be addressed through advances in artificial intelligence (AI), particularly in computer vision (CV). Early attempts to automate visual information processing began over three decades ago with one of the first convolutional neural networks recognizing handwritten zip codes on postage envelopes (*5*). Since then, CV algorithms have significantly progressed, finding numerous applications in scientific research. Deep learning models have been used to identify two-dimensional materials in microscopic images (*6*), characterize mineral composition in scanning electron microscopy (SEM) samples (*7*), and determine nanotube chirality from transmission electron microscopy images (*8*). Here, we report on using a deep learning algorithm to automatically recognize and track individual carbon nanotubes during growth, captured in polarization microscopy video sequences.

**Materials and methods**

*CNT growth*

The experimental setup was previously described in references (*4, 9*). In short, horizontally-aligned carbon nanotubes (HA-CNTs) were synthesized inside a miniature chemical vapor deposition (CVD) cell with an optical window (Linkam TS1500). ST-cut quartz and iron nanoparticles served as substrate and catalyst, respectively. Ethanol and argon were, respectively, used as carbon precursor and carrier gas. Oxygen and water sensors monitored gas-phase contaminants at the outlet line.

*In situ microscopy*

Nanotube growth was imaged *in situ* using a custom-built optical setup for homodyne polarization microscopy. A supercontinuum source (Fianium SC-400-4, 2 ps pulses, 40 MHz, spectral range 400-2000 nm) provided white light excitation across the visible spectrum. Two crossed polarizers were employed, with a polarizer and analyzer used to enhance the scattered field from the nanotubes relative to the stronger reflected field from the substrate. A low-pass optical filter with a cutoff wavelength of 700 nm was placed after the analyzer to filter out blackbody radiation generated by the heating crucible. A long-distance objective (Nikon Plan Fluor ELWD 20x 0.45 C L) was used for illumination and collection. Growth process videos were captured using a digital camera (Hamamatsu c11440 ORCA-Flash4.0 LT) with maximum acquisition rate up to 40 frames per second (fps).

*Video processing*

To extract kinetic data from each nanotube, the raw videos were first processed to enhance contrast: the detailed processing methodology is provided in the supporting information. In short, the initial frame rate, typically between 40 and 25 fps, was reduced by averaging frames to one frame per second. This operation boosts further analysis and improve signal-to-noise ratio without compromising kinetic data quality, as the nanotube growth rate is typically in the order of 1 µm/s and the localization precision of the optical setup is 0.33 µm. Frames were then aligned using a template matching algorithm from the OpenCV Python library. Shade correction was applied to compensate for uneven illumination caused by the optics. Residual noise was reduced using fast Fourier transform (FFT) band-pass filtering, and object edge contrast was enhanced using Gaussian difference filtering. Finally, image contrast was optimized through histogram equalization (*10*).

*Deep learning model*

The deep learning model and its training are detailed in the main text. In short, an image recognition system was implemented using the PyTorch library (*11*). For recognizing nanotubes in videos, we employed the Mask-RCNN architecture (*12*) with the ResNet-50 neural network as a backbone (*13*), pre-trained on the Microsoft COCO-2017 dataset (*14, 15*). Training was conducted on graphics processing units (GPUs) provided by Google Colab (*16*) over 150 epochs with a learning rate of 0.005, decreasing by 10 % every 5 epochs. A weight decay of 0.0005 with momentum of 0.9 was used.

**Results**

*In-situ* videos of CNT growth obtained using HPM contain extensive kinetic information. However, the low contrast of raw videos (see Figure 1a) is inadequate for kinetic assessment, necessitating raw video processing for statistical data acquisition on nanotube growth kinetics. A rolling-frame method (also called differential-treatment) was developed to enhance contrast (see details in Supplementary Information). This approach, which is a variation of the shade correction commonly used in image processing, involves subtracting a background snapshot containing the illumination profile from the image (*17, 18*). Typically, the first frame of a sequence is used as the background. However, using a frame with a fixed delay time (chosen between 5 and 30 seconds) significantly increases contrast, by up to an order of magnitude (see Figure 1b and Figure S4 VIDEO S01). The method is termed "differential" or "rolling frame" shade correction due to the rolling of the background and processed frames.

In such differential videos, the length of the nanotube segment is proportional to its instantaneous rate, as described in Equation S5, which adds useful information to the video. Additionally, differential videos capture other processes causing local changes in optical absorption. For instance, if the nanotube structure (also called helicity or chirality) changes during growth, this manifests as a second segment moving synchronously with the first one: the new chirality appears either as a bright segment if it has lower optical absorption, or as a dark segment otherwise. If the nanotube switches from growth to shrinkage, it appears as a single bright segment moving backward, corresponding to lower optical absorption (*19*).

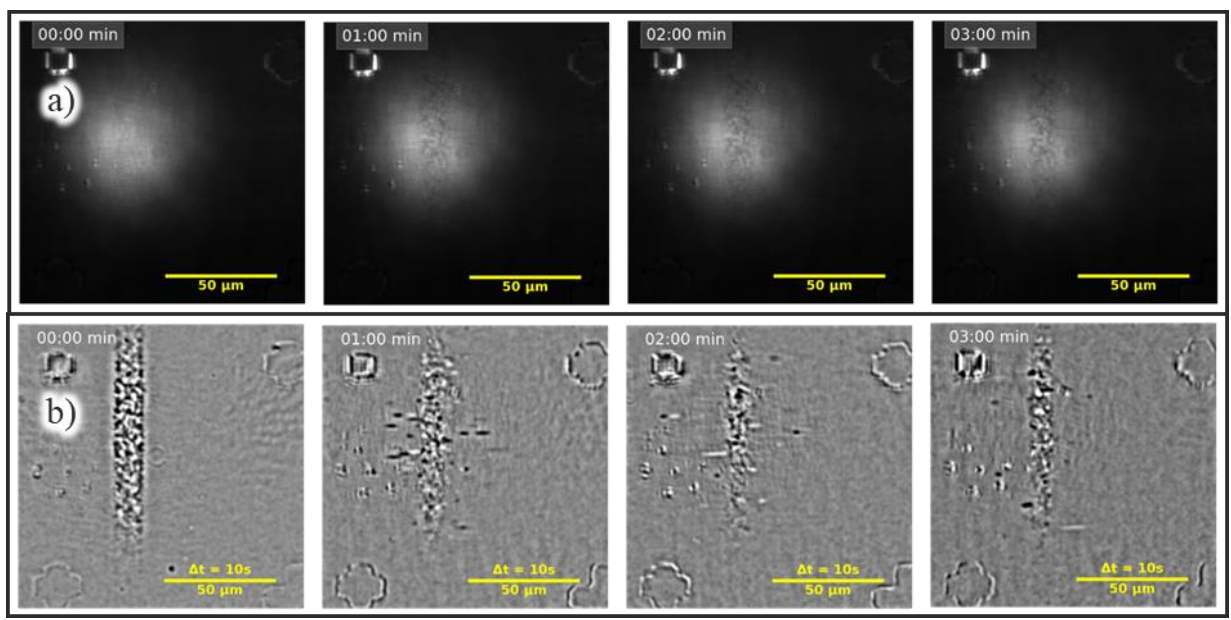

Figure 1. Snapshots from a) raw video of nanotube growth, b) fully processed video treated with differential shading correction (with a delay time of 10 s).

We developed a deep learning model to recognize and track both growing nanotubes (dark segments) and structural changes (bright segments) in such differential videos. The model was also trained at recognizing optical marks and catalyst lines (see Figure 2) (*19*). Kinetic data extraction proceeded in the following steps:

1. Object recognition
2. Tracking of recognized objects
3. Verification of tracking
4. Kinetic curve extraction and analysis

The initial stage of the process utilizes the Mask-RCNN neural network implemented in PyTorch Python library (*11*). This architecture integrates several neural networks in a single pipeline (see Figure 2). The image is first processed by the backbone network to produce a feature map. Region of interest (ROI) proposal and prediction networks then localize boundary boxes of the objects and classify them. Finally, a fully connected convolutional network generates a super pixel mask of the recognized objects (*20*).

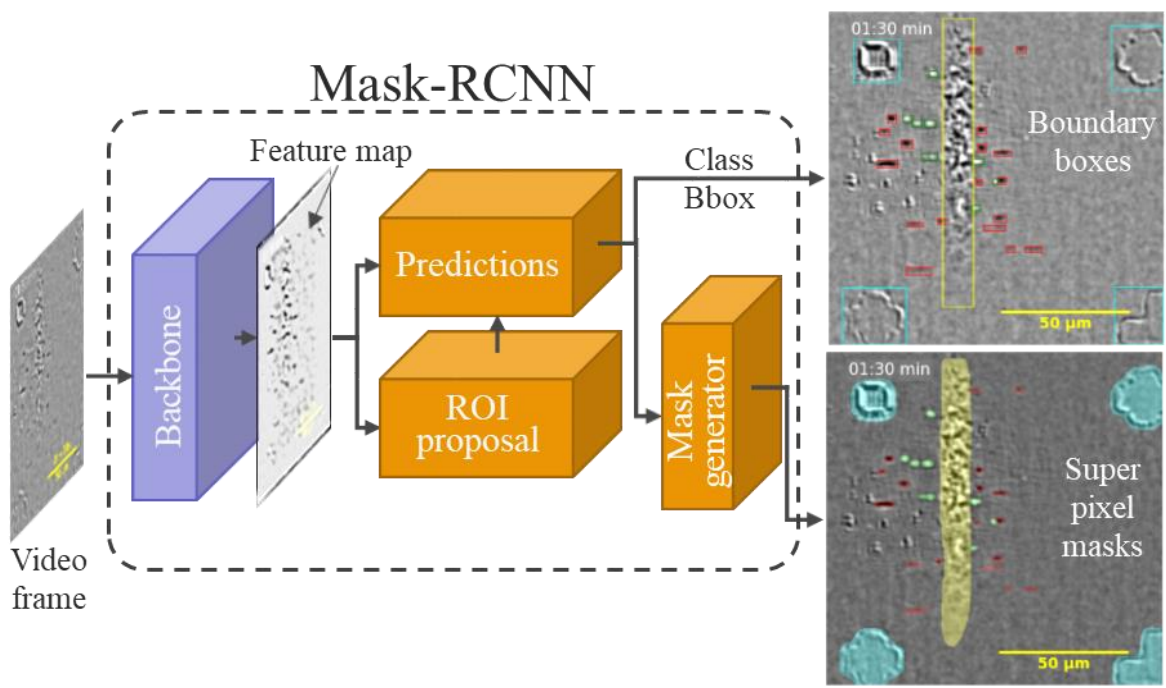

Figure 2. Schematic of object recognition system based on Mask-RCNN. Colors on the right inset highlights growing nanotube segments (red), structural changes (green), catalyst line (yellow), optical marks (blue).

The training dataset consisted of 580 manually labeled images from various videos. During training, the dataset was split into a training set of 550 images and a validation set of 30 images. In image segmentation, this data amount does not hinder model quality, and subsets as small as 30 images have been reported sufficient in some cases (*21, 22*). Given the varying quality of our videos, augmentations were applied to account for possible variations in image brightness, contrast, and nanotube localization (Figure S5). This approach not only expanded the dataset without additional manual labeling but also increased the robustness and stability of the trained model (*23*) (Table S1).

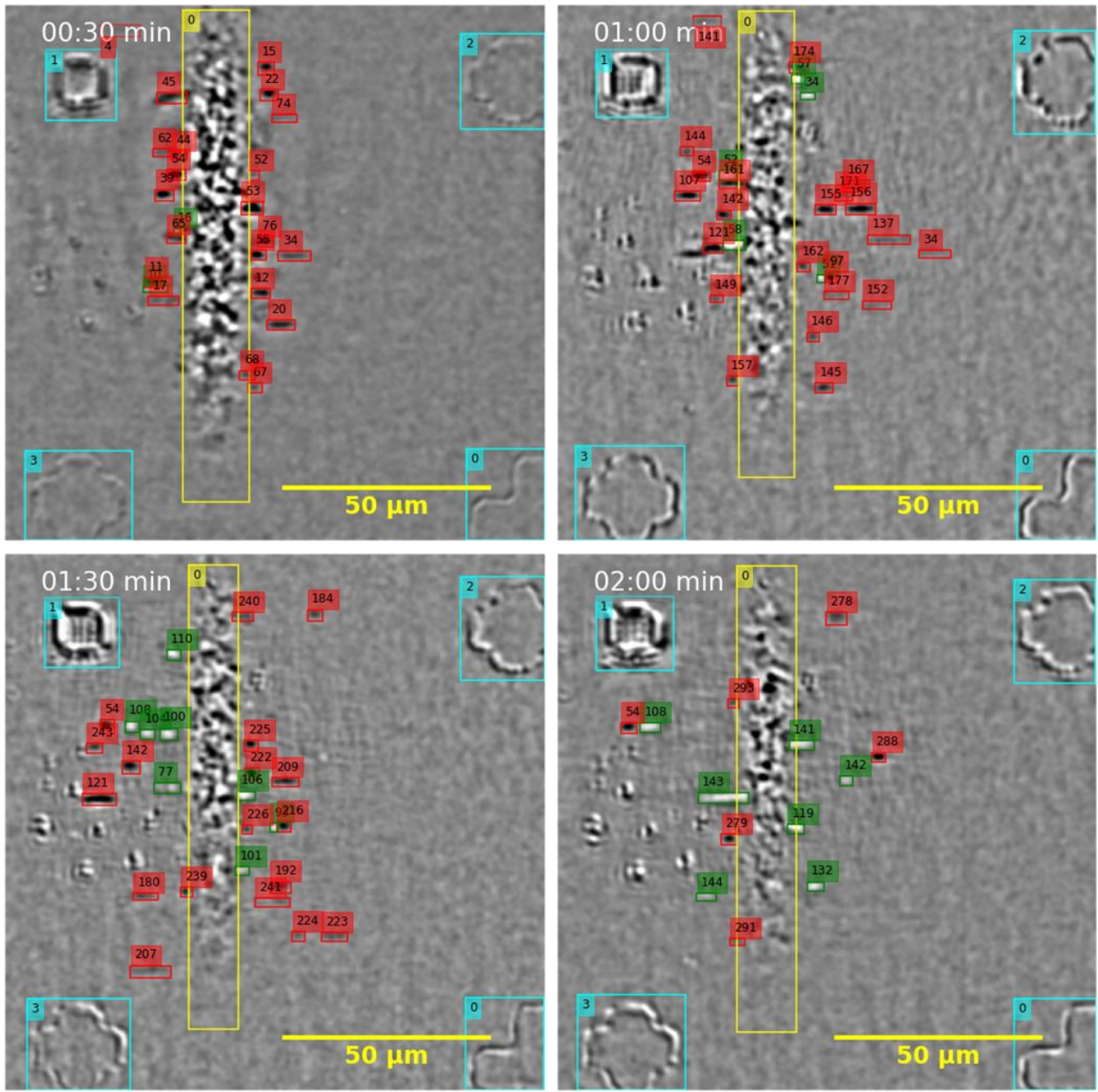

Figure 3. Snapshots from the in-situ video with boundary boxes around recognized objects and their numbers in the dataset.

The fully trained model detected segments corresponding to nanotube growth and structural changes, as well as optical marks and catalyst lines (Figure 3). This process was conducted frame by frame through the video. To extract the growth kinetics of each nanotube, the as-recognized objects were tracked using the Hungarian method (*24*) and Kalman filter (or Linear Quadratic Estimation) (*25*), which are widely used for object tracking (*26*). The Hungarian algorithm matched masks across successive frames. However, some video frames are unrecognizable due to imaging artifacts, illumination instability, or uncompensated vibrations. Following the initial tracking stage, all segments were grouped into clusters of varying sizes, corresponding to objects recognized across consecutive frames. The clusters were then subjected to a Kalman filter to merge segments corresponding to the same nanotube (See VIDEO S03).

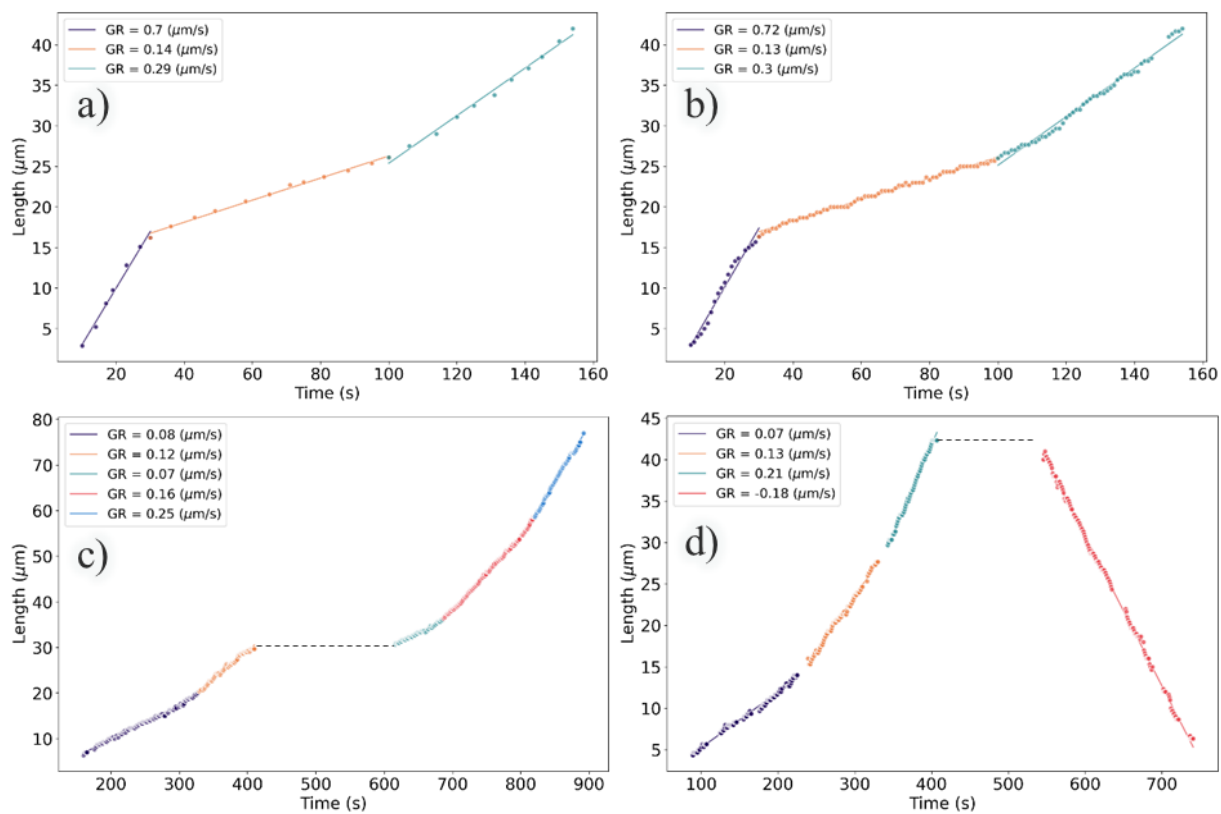

Figure 4. Kinetic curves of individual nanotubes. a) and b) correspond to the same nanotube measured manually and using automatic recognition system, respectively. c) and d) correspond to nanotubes grown with a pause between two stages of growth and between growth and etching, respectively. The points stand for measured data, lines correspond to a linear fit of the segments of constant growth (slopes are given in inset). Black dashed lines show the pauses during the nanotube growth.

The information about tracked segments is entered into tables for final manual verification and labeling of events. This manual step remains essential due to the complexity of nanotube kinetics, which involves switches between growth, pauses, shrinkage, and structure change during growth (*19*). Pauses (see Figure 4 c, d) cannot be efficiently traced by the Hungarian method or Kalman filter, necessitating manual verification to ensure correct assignment of newly grown segments to the same or another close nanotube (see Figure 4). Despite this final manual check, automating recognition and tracking steps has accelerated kinetic extraction, significantly enhanced the time resolution of kinetic curves (Figure 4 (b-d)), and increased the repeatability of kinetic measurements. A comparison of manually measured and DL model-extracted results is shown in Figure 4 (a,b).

Figure 5 compares the main kinetic parameters extracted manually and using the AI-assisted method from *in-situ* videos of nanotube growth performed at the same growth temperature and ethanol partial pressure. In figure 5a, the lifetime and segment growth duration represent the time during which the nanotube grew at a constant rate (see dots clusters of the same color in Figure 4). Linearly grown nanotubes exhibit constant growth rates from start to finish, observed in about half of the cases. The other half show one or more stochastic rate changes, or even growth-shrinkage switches separated by pauses (*19*). The dependence of growth rate on the growth duration is shown in Figure 5b. The proximity of centers, intersection of standard deviations, and similarity between distributions of kinetic parameters extracted manually and by AI-assisted tracking demonstrate the consistency of both approaches and their applicability for linear and non-linear growth analysis. The difference in the numbers of plotted instances in Figure 5 stems from manually analyzed videos used for model training, which cannot be used for AI recognition to avoid bias. The AI-extracted data in Figure 5 represents a small fraction of all data acquired using this method, forming the foundation for developing a statistically supported model (*27*). On average, extracting kinetic data from a single video takes approximately 6 hours for a time resolution of 5 s using the manual method, compared to just 2 hours for a time resolution of 1 s with the deep learning model, meaning that the throughput was increased by about a factor 15 (see Figure 4a, b).

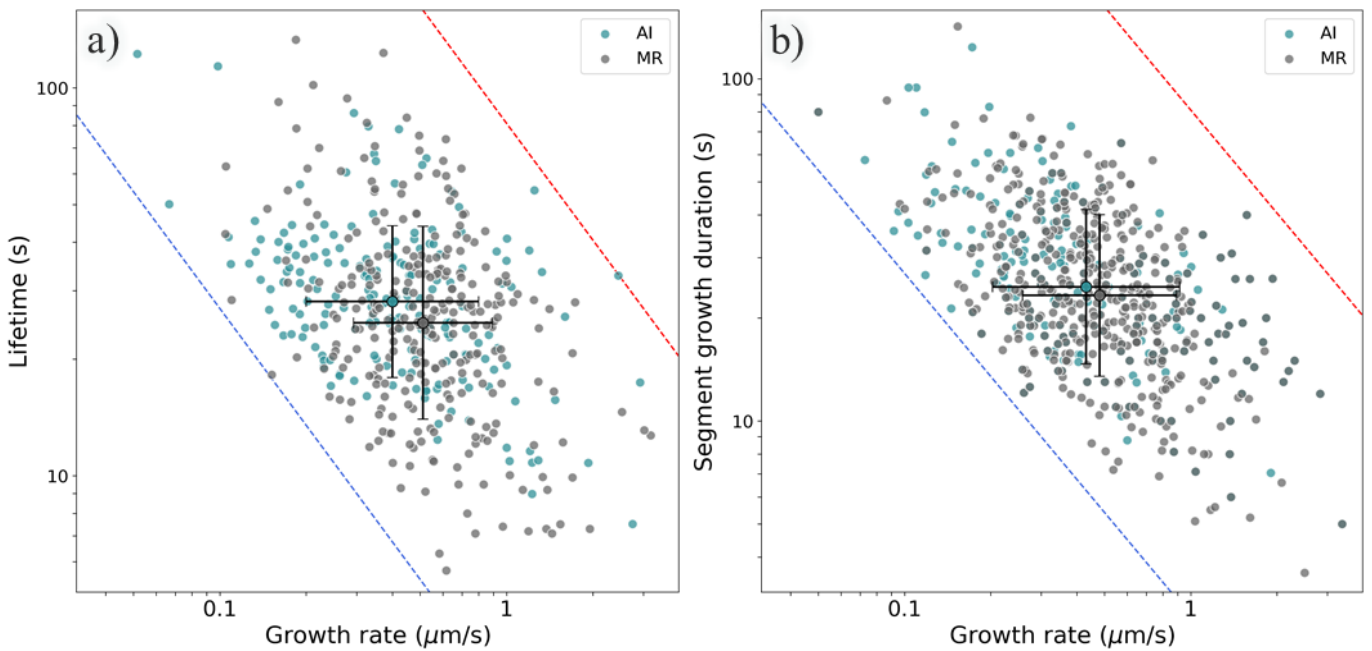

Figure 5. Plots of growth duration (lifetime) versus growth rate for individual nanotubes: a) nanotubes with constant growth rate and b) nanotubes with growth rate changes. Green and grey dots correspond to the data extracted using automated system and manual measurements, respectively. The large dots show the mean of the distributions and whiskers stands for the one standard deviations along the corresponding axes.

**Conclusion**

We have identified several critical steps in developing a DL model for extracting kinetic data from *in-situ* microscopy videos of individual nano-objects. A specific challenge is the low signal from nano-objects and fast kinetics on the video time scale. Applying differential treatment not only drastically improves contrast but also enhances human operator understanding of ongoing processes. This is crucial for manual recognition and tracking and for properly labeling the dataset used to train the DL model, allowing comparison of data obtained by both methods to validate their coherence. Properly designed training datasets and augmentations improve model robustness and stability without increasing training data volume.

Despite significant advancements, there is room for improvement in processing algorithms. In particular, manual verification and labeling of tracked nanotubes, particularly for complex cases, remains essential and is currently the most time-consuming step in the process (*19*). Currently, brightness and contrast adjustments rely on empirical hyper-parameters, which could benefit from self-tuning or non-parametric algorithms. The same applies to FFT and Gaussian filtering parameters, which are presently constant. Additionally, the AI-based algorithm for kinetic extraction from differentially treated videos can be refined, regardless of the performance comparable to the current state-of-the-art models (*31*). Expanding the dataset to include more diverse video quality and nanotube localization can lead to more stable models (*32*). Moreover, replacing computationally demanding models like Mask RCNN with smaller models could enhance overall performance, although potentially reducing recognition accuracy.

Automating recognition and tracking is key to high-throughput video analysis, which is crucial for understanding and modeling complex nanoscale processes. The approach demonstrated here is versatile enough to be applied to recognizing and tracking individual nanostructures in other types of videos, such as environmental transmission electron microscopy of carbon nanotubes (*28, 29*) or environmental scanning electron microscopy of graphene growth and etching (*30*).

**Supplementary files:**

Supplementary Information (pdf): Expanded description of the differential video processing, Comprehensive explanation of model training process, evaluation of different models, and description of tracking process

Supplementary video files:

There are two groups of videos demonstrating different processing, recognition and tracking steps, numbered Video_XX SYY, where XX stands for the number of videos and YY to the processing step.

Video_01 – *In-situ* videos demonstrating different processing, recognition and tracking steps (this video was not used for model training, kinetic were extracted using AI assisted system from the video S03)

Video_02 – *In-situ* video demonstrating different treatment steps (have been used for the model training, kinetic were extracted manually)

S01 – Video showing raw, fixed frame and rolling frame *in-situ* sequences

S02 – Differential video after recognition and Hungarian algorithm tracking (masks on the left, boundary boxes on the right)

S03 – Finalized differential video after recognition, tracking by Hungarian algorithm, Kalman filter and manual verification (masks on the left, boundary boxes on the right)

**Acknowledgments:** The authors acknowledge the support of the *Agence Nationale de la Recherche* (grant ANR-20-CE09-0007-01).

**Author contributions:** V.J. designed and coordinated the research. S.T. prepared the catalyst samples and performed SEM characterization. V.P. performed the *in situ* optical imaging, developed the protocols of data treatment and analyzed the experimental data.